# Observation of linear magnetoelectric effect in a Dirac magnon antiferromagnet $Cu_3TeO_6$

Aga Shahee,[1,2] Kyongjun Yoo,[1] B. Koteswararao,[1,3] N. V. Ter-Oganessian[4*] and Kee Hoon Kim[1,5†]

[1]*Center for Novel States of Complex Materials Research, Department of Physics and Astronomy, Seoul National University, Seoul 151-747, Republic of Korea*

[2]*Centre for Interdisciplinary Research & Innovations, University of Kashmir, Hazratbal, Srinagar, Jammu and Kashmir 190006, India*

[3]*Department of Physics, Indian Institute of Technology Tirupati, Tirupati 517 506, India*

[4]*Institute of Physics, Southern Federal University, Rostov-on-Don 344090, Russia*

[5]*Institute of Applied Physics, Department of Physics and Astronomy, Seoul National University, Seoul 151-747, Republic of Korea*

## ABSTRACT

$Cu_3TeO_6$, a three-dimensional antiferromagnet forming a unique spin-web lattice of spin-1/2 $Cu^{2+}$ ions below the Néel temperature $T_N \approx 62$ K, has recently been found to exhibit topological Dirac or nodal magnon dispersion. In this study, we report the discovery of the linear magnetoelectric (ME) effects in $Cu_3TeO_6$ below $T_N$. Our pyroelectric current measurements at a constant magnetic field ($H$) reveal a linear increase of electric polarization ($P$) with $H$ for both $P$ // $H$ and $P \perp H$ configurations; a maximum $P$ // [110] = 20 μC/m$^2$ is obtained at $\mu_0 H$ //$[1\bar{1}0]$ = 14 T, corresponding to a linear ME coefficient 1.8 ps/m. Magnetic point group analysis and Monte-Carlo simulations confirm that finite linear ME coefficients are allowed in the off-diagonal and diagonal ME tensor components, consistent with the magnetic point group of $\bar{3}'$. As the parity-time symmetry can be broken in the presence of $H$ or electric field $E$ in the linear ME materials, we envisage that $Cu_3TeO_6$ should exhibit a $H$- or $E$-induced transformation in the topological magnon dispersion from a Dirac point/nodal line type into two Weyl point types.

**Corresponding Author -** Kee Hoon Kim

Postal address : #56-427, Seoul National University, 1 Guanak-ro, Guanak-gu, Seoul, South Korea.

E-mail: khkim@phya.snu.ac.kr Tel: +82-2-880-9068

## INTRODUCTION

In the linear magnetoelectric (ME) materials, electric polarization (*P*) can be linearly induced by magnetic field (*H*) or magnetization (*M*) can be linearly induced by electric field (*E*). The linear ME effect was first discovered experimentally in an antiferromagnet $Cr_2O_3$ by Astrov, [1] after the theoretical predictions first by Landau and Lifshitz [2] and later by Dzyaloshinskii. [3] The linear ME effect occurs when a material breaks space inversion (*I*) and time-reversal (*t*) symmetry separately but conserves the simultaneous operation of $I \otimes t$.

Since the first discovery, numerous linear ME materials have been discovered and studied, particularly in the 1960-1970s [4,5]. Furthermore, in the early 2000s, research interests on the ME effects were reignited by the discoveries of large non-linear ME effects in several multiferroic materials [6–8]. Over the last two decades, research on the multiferroics has been very active, leading to the discovery of numerous emergent multiferroics and new mechanisms for generating ME coupling [9,10]. Very recently, even 2D van der Waals materials have been found to exhibit multiferroic behavior [11]. Moreover, a topologically protected switching mechanism has been recently identified in a multiferroic $GdMn_2O_5$ (Ponet et al., 2022). Additionally, while $BiFeO_3$ films have been studied for a long time, Shin et al., [13] successfully grew high-quality thin films of $Co_2Z$ hexaferrites exhibiting ME behavior at room temperature. Thus, scientific research on multiferroic and magnetoelectric materials seems to be still actively performed.

The ME materials possess unique properties that enable cross-coupling between magnetic and electrical properties, making them increasingly valuable for various next-generation electronic devices. Examples include highly sensitive *H* sensors with pico-tesla sensitivity [14] and low-power non-volatile memories (Bibes and Barthélémy, 2008). Moreover, ME and/or multiferroic materials can be utilized as tunable filters, antennas, and resonators in millimetre-wave devices for wireless communication and radar [16], ME logic circuits, which operate at a low voltage of ~100 mV, provide an energy-efficient alternative to traditional CMOS technology [17], and flexible ME nanogenerators for energy harvesting [18]. Thus, recent research has demonstrated the potential of ME materials to offer promising alternatives to the conventional technologies, making them highly attractive for applications.

In addition to research efforts aimed at finding practical applications for ME materials, studying the ME effects in magnetic insulators can be useful for understanding their basic properties. For example, the ME tensor characteristics of a magnetic insulator can serve as an effective means of determining its magnetic point group when neutron scattering fails to determine magnetic symmetry unambiguously. Moreover, as found in this work, the ME effect can also be utilized to manipulate the topological magnon properties of a magnetic insulator.

$Cu_3TeO_6$ belongs to corundum related compounds of $A_3TeO_6$ ($A$ = Mn, Co, Ni, and Cu) family, which has recently attracted scientific attention due to its each members unique properties. Although they nominally have the same chemical formula, each member of $A_3TeO_6$ family exhibits different structural and physical properties based on the type of ions at the $A$-site. $Mn_3TeO_6$, for instance, adopts a trigonal structure ($R\overline{3}$ space group) and exhibits a complex incommensurate magnetic structure with multiferroic properties. [19,20] while $Co_3TeO_6$ crystallizes in a monoclinic structure ($C2/c$ space group) and exhibits $H$-induced electric polarization ($P$). [21,22] Moreover, $Ni_3TeO_6$ has a trigonal structure ($R3$ polar space group) and exhibits $P$ in the spin-flop phase induced by $H$ above 8 tesla (T). [23,24] On the other hand, $Cu_3TeO_6$ is known to crystalize in a cubic lattice structure of high symmetry ($Ia\overline{3}$ centrosymmetric space group) without any report on its electrical properties. It forms a unique bixbyite-type structure, in which 24 $Cu^{2+}$, 8 $Te^{6+}$, and 48 $O^{2-}$ ions are contained in a unit cell, forming a network of 8 hexagons composed of 6 $Cu^{2+}$ ions and a central $Te^{6+}$ ion (See, Fig. 1 (a)). [25,26]

Upon the temperature being lowered, $Cu_3TeO_6$ undergoes a paramagnetic to an antiferromagnetic (AFM) spin ordering at the Néel temperature ($T_N$) of 62 K. This leads to the stabilization of a unique three-dimensional spin-web lattice of $Cu^{2+}$ ions with $S$ = ½ spins. [25,27–29] An earlier report based on neutron powder diffraction and torque magnetometry measurements suggest the appearance of either a collinear AFM order or a slightly canted AFM order with the dominant spin moment along one of [±1±1±1] directions, [25] forming a trigonal symmetry in the spin system. [25] Recent theoretically predictions by Li *et al.* [30] suggest the existence of Dirac and the nodal line magnons for the collinear and spin canted state of $Cu_3TeO_6$, respectively. This was subsequently confirmed by Bao *et al.* [31] and Yao *et al.* [32] , who reported the spin excitation spectra of $Cu_3TeO_6$ from the inelastic neutron scattering (INS) measurements. These experiments provided experimental evidence for the novel topological Dirac/nodal line magnons. However, the resolution of the magnon spectra from those two experiments was not sufficient to distinguish possible nodal lines from the Dirac point.

According to the theoretical prediction using the first-principles calculations and a linear-response theory by Wang *et al.,* [33] the Dirac point (or the Nodal line) in the magnon dispersion is coined to the absence (or presence) of a spin canting in the AFM state. The spin canting in $Cu_3TeO_6$ is most likely induced by the Dzyaloshinskii-Moriya interactions (DMI). These topological Dirac point and nodal line in the magnon dispersion suggest that $I \otimes t$ symmetry is preserved without any biased $E$ and $H$, thus allowing the linear ME effect consistent with a magnetic point group. However, applied $H$ and $E$ for the ME poling may lead to the eventual stabilization of one type of ME domain and emergence of the transversal electric polarization. This would imply that the $I \otimes t$ symmetry is broken by the application of $H$ or $E$. In this case, the magnon bands are expected to carry non-zero Berry curvature/Chern numbers, which may transfrom each Dirac point/nodal line into two Weyl points with opposite charges. [34] Therefore, investigating the ME effects in $Cu_3TeO_6$ can further our understanding of characteristic magnon dispersion at zero and finite $H$ or $E$.

In this letter, we report the discovery of both off-diagonal and diagonal linear ME coupling in $Cu_3TeO_6$ below $T_N$ = 62 K. Our findings indicates that the electric polarization ($P$//[110]) exhibits a linear increase under a magnetic field ($H$//$[1\bar{1}0]$) up to 14 T with the configuration of $P \perp H$ . We conducted a combined symmetry analysis and Monte-Carlo calculations to confirm that the observed linear ME effect with both off-diagonal and diagonal components can be attributed to the magnetic point group $\bar{3}'$.

## EXPERIMENTAL METHODS

Single crystals of $Cu_3TeO_6$ were grown by the flux method. First, a polycrystalline $Cu_3TeO_6$ powder was prepared using high purity CuO (>99.99 %) and $TeO_2$ in molar ratio of 3:1 via the solid-state reaction method. The powder was calcined at 600 °C for 12 hours twice and sintered at 850 °C for 24 hours. Then single crystals were grown using a mixture of polycrystalline $Cu_3TeO_6$ and $PbO/TeO_2$ flux (a molar ratio 1:1) at a mass ratio of 1:2. The mixture was heated at 950 °C for 12 h and then cooled to 700 °C at a rate of 1.25 °C/h. Single crystals, ~1 $mm^3$ size and cubic in shape, were mechanically separated from the flux. The lattice structure of the grown $Cu_3TeO_6$ single crystals was characterized using an X-ray diffractometer (Empyrian™, Malverin Pananalytycal).

To conduct electrical measurements, the single crystals were shaped into a thin plate form along the [100] and [110] directions, and electrodes were made with silver epoxy on both sides of the plate. Magnetic properties were investigated using a vibrating sample magnetometer (VSM). Pyroelectric current ($I_p$) and dielectric constant (ε) were measured using an electrometer (Keithley, KE617) and a capacitance bridge (Andeen-Hagerling, AH2550A), respectively. These measurements were performed in a temperature ($T$) range from 2 to 75 K using a PPMS™ (Quantum Design, 9 T) or a Janis cryostat equipped with a 14 T superconducting magnet (Cryogenic$^{LTD}$). Prior to $I_p$ measurements, the sample was first cooled down from 70 to 2 K under applications of an electric field of 1.2 MV/m and magnetic fields of 0, 3, 9, 12, and 14 T. After cooling, the electric field was removed and both sides of the electrode were short-circuited until stray charges were fully nullified. The $I_p$ at zero electric field was recorded during temperature-warming mode under applications of constant magnetic fields. Electric polarization ($P$) was obtained by integrating $I_p$ over time. Furthermore, for all the configurations, it has been confirmed that the sign of the $P$ is reversed by reversing the direction of a poling electric field.

## RESULTS AND DISCUSSIONS

### *Magnetic and electrical properties*

The lattice structure of $Cu_3TeO_6$ and the arrangement of its magnetic ions ($Cu^{2+}$) are shown in Figs. 1 (a) and (b), respectively. $Cu_3TeO_6$ crystalizes in a bixbyite-type high-symmetric cubic structure ($Ia\overline{3}$ space group) with 12 $Cu^{2+}$ (S= ½) ions per primitive cell and with a lattice parameter of $a$ = 9.537 Å at room temperature. [26] Despite its simple structure, $Cu_3TeO_6$ exhibits a spin-web type antiferromagnetic ordering with topological magnon dispersion. The spin-lattice is composed of nearly coplanar hexagons formed by six $Cu^{2+}$ (S= ½) ions with normal along the [±1±1±1] directions. There are four different orientations of hexagons with normal along [111], [11-1], [1-11], or [-111], and each $Cu^{2+}$ ion is shared by two hexagons, forming an unusual hyper kagome network (see, Fig. 1 (b)). A nearly straight Cu-O-O-Cu path can be observed between the 9$^{th}$ nearest neighbors. Recent INS [32] measurements and first-principles calculations [33] have shown that the strength of the 9$^{th}$ nearest-neighbour exchange interaction, which is antiferromagnetic, is comparable to the first nearest neighbor interaction, while the second and other nearest-neighbour exchange interactions are rather weak.

X-ray diffraction (XRD) data measured in the $Cu_3TeO_6$ single crystal with {100} plane are shown in Fig. 1 (a). The data confirm that the $Cu_3TeO_6$ single crystal was grown without chemical impurities. Since the $Cu_3TeO_6$ has a cubic structure, the crystals have a clear rectangular shape with a size of ~ 1 mm$^3$, as seen in the inset of Fig. 1(c). Figure 2(a) displays the magnetic susceptibility ($\chi$) data measured at $H$ = 1 kOe along the cubic directions [100], [110], and [111] after zero-field cooling. The $\chi$ curves exhibit a downturn at $T_N$ = 62 K along all directions, indicating the onset of a long-range AFM ordering below $T_N$. The $\chi$ curves are similar for each direction but show a minimum value along the [111] direction. [25,28], suggesting that the [111] direction is the easy axis, although the magnetic anisotropy is quite weak. These results are consistent with earlier neutron scattering data, [25] that suggest the spins are oriented parallel to the [111] axis in the AFM phase and the magnetic anisotropy is quite weak. This behavior is also qualitatively consistent with magnetic data reported by other groups. [28,29,35,36]

The Curie-Weiss fit of the inverse $\chi$ (Fig. 2(b)) shows the Curie-Weiss temperature $\theta_{CW}$ = -160 K, whose absolute value is much higher than $T_N$. This $\theta_{CW}$ = -160 K is consistent with previously reported values ranging from -134 K to -213 K. [25,29,31,32] and indicates that the magnetic interactions are predominantly AFM. The value of the frustration index ($f = |\theta_{CW}|/T_N$) of 2.58 in our case and 2.23 – 3.55 in other previous reports suggests that the $Cu_3TeO_6$ is moderately frustrated. [25,29,31,32] It has been reported that except for the strong first "$J_1$", & ninth "$J_9$" and the moderate second "$J_2$", fourth "$J_4$", & tenth "$J_{10}$" exchange interactions, all other exchange interactions up to $J_{20}$ are rather weak. The strong $J_1$ and $J_9$ are compatible with the collinear AFM ground state, while the modest frustration possibly comes from the incompatibility of the sizeable $J_2$, $J_4$, and $J_{10}$ exchange interactions with the collinear AFM state. [33] Besides, it is further found that $\chi$ starts to deviate from the Curie-Weiss law below ~126 K possibly indicates the onset of short-range spin order. [32]

We also measured the $H$-dependent magnetization ($M$) along three crystallographic directions ([100], [110] & [111]) at 5 K (Fig. 2(c,d)). The $M$-$H$ curve along the [111] direction shows a slight curve between -1.5 T and 1.5 T, whereas it is almost linear for the [100] and [110] directions. [25] This change in the slope of the $M$-$H$ curve along the [111] direction is due to the presence of four types of domains, each with the spins oriented along one of the body diagonal [111], $[\bar{1}11]$, $[\bar{1}\bar{1}1]$, $[\bar{1}\bar{1}\bar{1}]$ directions of the cubic unit cell, which then yield their $H$ -induced rotation and alignment. [25] The magnetization $M$ increases almost linearly up to ~0.086 $\mu_B$/f.u at 9 T, which is still much lower than the expected saturation value of $M_s$ = 3 $\mu_B$/f.u. This confirms that the AFM spin configuration is maintained at least up to 9 T.

To investigate ME coupling in $Cu_3TeO_6$, we measured the pyroelectric currents ($I_p$) and dielectric constant $\varepsilon$ at selected $H$ in a transverse configuration, i.e., $I_p$ and $\varepsilon$ //[110] and $H$//[1-10], as summarized in Fig. 3 (a–d). Without the application of $H$, no observable $I_p$ is measured, as shown in Fig. 3 (a). However, upon application of $H$ = 3 T, $I_p$ at low temperature starts to decrease linearly above 20 K and shows a sharp dip near $T_N$ = 62 K. As $H$ increases beyond 3 T, the linear decrease and dip feature in $I_p$ are progressively enhanced up to a maximum applicable field of $H$ = 14 T, implying that $P$ is continually enhanced with increasing $H$. The sign of $I_p$ was reversed by a negative electric poling (-$E_P$) (data not shown here), which indicates that $P$ is intrinsically induced by the broken inversion symmetry under $H$.

Figure 3(b) summarizes the temperature-dependence of the $P$ obtained by integrating the $I_p$ with time. For all applied $H$, $P$ is nearly constant below 20 K, decreases above 20 K approximately as $\sqrt{T_N - T}$, and finally drops to zero at $T_N$. The reported temperature dependence of the staggered moment obtained from the neutron diffraction measurement [25] is similar to that of the $P(T)$ curves. This implies that the evolution of the magnetic order parameter in $Cu_3TeO_6$ below $T_N$ is directly coupled to the evolution of $P$ under $H$. These results provide further evidence of the strong ME coupling in $Cu_3TeO_6$.

In Fig. 3(c), we summarize the resultant $P$ value at 2 K at each constant $H$. The figure shows that the $P$ value increases nearly linear with $H$ until it reaches the maximum value of 20 $\mu C/m^2$ at $\mu_0 H$ =14 T, which supports the linear ME coupling. We obtained linear ME coefficient $\alpha$ of 1.8 ps/m, which is lower than those observed in other important linear ME compounds such as $TbPO_4$ (730 ps/m, 1.5 K), [37] and $LiCoPO_4$ (30.6 ps/m, 4.2 K) [38]. However, the $\alpha$ value of ~1.8 ps/m is higher than those of $BiFeO_3$ (0.3 ps/m, 300 K), $GaFeO_3$ (0.06 ps/m, 300 K) [39], $(NH_4)_2[FeCl_5 \cdot (H_2O)]$ (1.2 ps/m, 1.5 K) [40], and $LiNiPO_4$ (1.7 ps/m, 20 K) [41] although it is comparable to those of $Cr_2O_3$ (2.8 ps/m, 300 K) [42], $NdCrTiO_5$ (-1.84 ps/m, 10 K) [43] and $Sm_2BaCuO_5$ (4.4 ps/m, 2 K) [44]. The relatively small ME effect of $Cu_3TeO_6$ might be due to the isotropic nature of $S$ = ½ spins and the small spin-orbit coupling. It's worth noting that this linear ME effect is quite unique in the $Cu^{2+}$ ($S$=1/2) systems since most of the linear ME effect have been discovered in large spin ($S$>1/2) systems with magnetic ions such as Cr, Fe, Mn, and Ni. Recently, in the family of $Cu^{2+}$ ($S$ = ½) system, we observed $H$-induced ferroelectricity in a kagome staircase compound $PbCu_3TeO_7$, whose origin is related to the type II multiferroicity coming from a spiral spin structure. [45]

Figure 3(d,e) summarizes the temperature dependence of the dielectric permittivity $\varepsilon$ for $E$ // [110] at several magnetic fields (i.e. 0, 3, 6, 9, 12 & 14 T) applied along the $H$ // [1-10] direction. At zero $H$, $\varepsilon$ exhibits a weak drop at $T_N$ = 62 K. When $H$=3 T is applied, there is a sharp jump of $\varepsilon$ at $T_N$ and the magnitude of this jump increases greatly with increasing $H$. It is worth noting that $\varepsilon$ at 2 K increases tremendously from ~13.4 at 0 T to 29.8 at 14 T. Moreover, the dielectric loss also shows a large change under $H$ (not shown here). The magneto-dielectric (MD) effect, calculated using the formula, $\mathrm{MD}\ (H) = \frac{\varepsilon'\ (H) - \varepsilon'\ (0)}{\varepsilon'\ (0)} \times 100\%$, is summarized in Fig. 3(f) as a function of $T$ at different $H$. This shows that $Cu_3TeO_6$ exhibits a positive, giant MD response below $T_N$ with maximum MD of 123% at 2 K at $H$ = 14 T. The observation of the giant MD effect in $Cu_3TeO_6$ directly reflects the presence of large modulations in the lattice degrees of freedom under an applied $H$, thereby supporting the intrinsic nature of the magnetoelectric effect.

To investigate the possibility of a ME response induced by $H$ in other crystallographic directions and configurations, we conducted measurements of the $I_p$ curves in different $H$ and $P$ directions. The $P$ curves obtained by integrating the $I_p$ data over time are presented in Fig. 4. These $P$ curves show that the polarization for all configurations emerges at $T_N$, at which $\varepsilon$ exhibits an anomaly for all the investigated configurations. Further, the $P(T)$ curves show the same line shapes but different values for different directions. At 2 K and 9 T, $P$ values for the ($P$//[110], $H$//[1-10]), ($P$//[110], $H$//[110]), ($P$//[100], $H$//[010]), and ($P$//[100], $H$//[100]) configurations are 13.3 μC/m$^2$, 1.4 μC/m$^2$, 7.4 μC/m$^2$, and 0.8 μC/m$^2$, respectively. These observations indicate that the emergence of $P$ with $H$ in this linear ME system is not restricted to only one highly symmetric lattice direction. Among them, the maximum ME coefficient of 1.8 ps/m is realized in the direction of $P$//[110] with $H$//[1-10]. It is worth noting that $P$ is an order of magnitude larger in the $H \perp P$ configurations than in the $H$ // $P$ configurations, indicating that the off-diagonal components of the linear ME tensor are dominant. These experimental results thus reveal that a particular form of the linear ME tensor associated with a specific magnetic point group should be responsible for the induced $P$ at a finite $H$ below $T_N$.

***The symmetry analysis and the classical Monte-Carlo simulations***

Next, we would like to discuss the possible mechanism of ME coupling in $Cu_3TeO_6$. For that, we performed symmetry analysis and conducted classical Monte-Carlo simulations using the exchange constants available from the literature data. [32,33] According to the neutron diffraction data, [25] the magnetic structure appearing below $T_N$ is described by the wavevector $\vec{k} = 0$. The magnetic representation $\Gamma_m$ for the 12 magnetic $Cu^{2+}$ ions in the primitive cell is split into

$$\Gamma_m = \Gamma^{1+} \oplus \Gamma^{1-} \oplus \Gamma^{2+3+} \oplus \Gamma^{2-3-} \oplus 5\Gamma^{4+} \oplus 5\Gamma^{4-},$$

where $\Gamma^{2+3+}$ and $\Gamma^{2-3-}$ are physically irreducible representations (IR). The AFM ordering observed by the neutron diffraction experiment is described by IR $\Gamma^{4-}$, which enters five times into $\Gamma_m$ and according to which five order parameters are transformed. Three of these order parameters $(a_1, a_2, a_3)$, $(b_1, b_2, b_3)$, and $(c_1, c_2, c_3)$ describe the collinear magnetic structure. The phase state of the form $(a, a, a)$ with all three order parameters being equal gives the direction of spins along the [111] direction. The proposed noncollinearity is described either by the remaining two order parameters transforming according to IR $\Gamma^{4-}$, which are necessarily condensed at the $\Gamma^{4-}$ instability at $T_N$, or by the order parameter $q$ described by IR $\Gamma^{1-}$, which is induced by $\Gamma^{4-}$ in an improper way due to the interaction

$$a_1 a_2 a_3 q.$$

The magnetoelectric interactions that result in *H*-induced electric polarization below $T_N$ are

$$I_1 = a_1 M_y P_z + a_2 M_z P_x + a_3 M_x P_y, \qquad (1)$$

$$I_2 = a_1 M_z P_y + a_2 M_x P_z + a_3 M_y P_x, \qquad (2)$$

where any of the five order parameters transformed according to $\Gamma^{4-}$ can replace $(a_1, a_2, a_3)$. As mentioned above, the magnetic structure belongs to the phase state $(a, a, a)$, which has eight domains (the spins can be oriented predominantly along one of the four spatial diagonals of the cubic cell). Magnetoelectric poling prior to the measurements partially lifts this degeneracy.

For the analysis of the ME behavior one can write the expansion of the thermodynamic potential as a function of the order parameters in the form

$$\begin{aligned}
\Phi &= \Phi_0 + \kappa_1(I_1 - I_2) + \kappa_2(I_1 + I_2) - \vec{M}\vec{H} - \vec{P}\vec{E} \\
&= \Phi_0 + \kappa_1\left[a_1\left(M_y P_z - M_z P_y\right) + a_2(M_z P_x - M_x P_z) + a_3\left(M_x P_y - M_y P_x\right)\right] \\
&+\kappa_2\left[a_1\left(M_y P_z + M_z P_y\right) + a_2(M_z P_x + M_x P_z) + a_3\left(M_x P_y + M_y P_x\right)\right] - \vec{M}\vec{H} - \vec{P}\vec{E}, \qquad (3)
\end{aligned}$$

with

$$\begin{aligned}
\Phi_0 &= \frac{A}{2}(a_1^2 + a_2^2 + a_3^2) + \frac{B_1}{4}(a_1^4 + a_2^4 + a_3^4) + \frac{B_2}{4}(a_1^2 + a_2^2 + a_3^2)^2 \\
&+ \frac{1}{2\varepsilon}\left(P_x^2 + P_y^2 + P_z^2\right) + \frac{1}{2\chi}\left(M_x^2 + M_y^2 + M_z^2\right),
\end{aligned}$$

where *A*, $B_1$, and $B_2$ are phenomenological coefficients, and $\varepsilon$ and $\chi$ are dielectric and magnetic susceptibilities, respectively.

According to our results, the $H$-induced polarization for the *P*//[110] and *H*//[1-10] geometry is much larger than for *P*//[110] and *H*//[110]. In the former case, the ME coefficient is $\kappa_1$, while in the latter, it is $\kappa_2$. This indicates that $\kappa_1$ is much larger than $\kappa_2$. In the case of *P*//[100] and *H*//[010], the ME coefficient is $\kappa_1 - \kappa_2$ and, therefore, should be close to the case of *P*//[110] and *H*//[1-10]. However, for the former case (i.e., the case of *P*//[100] and *H*//[010] in Fig. 4**c**), the *H*-induced polarization is about as twice lower than for the latter case (i.e., the case of *P*//[110] and *H*//[1-10] in Fig. 4**a**). In our view, this larger than-expected difference is the result of higher order ME interactions with respect to the AFM order parameter. Indeed, there are many additional third-order ME interactions in the AF order parameter, which are linear with respect to magnetization and electric polarization, e.g.,

$$a_1 a_2 a_3 \left(M_x P_x + M_y P_y + M_z P_z\right).$$

From our point of view, such interactions are responsible for the electric polarization induced by *H* in the *P*//[100] and *H*//[100] geometry, for which interactions (1) and (2) give zero ME effect.

To gain further insights into the microscopic interactions, we rewrote the magnetoelectric interactions (1) and (2) in terms of the spins of the ions in the primitive cell, and we analyzed and discussed the resulting numerous microscopic interactions in the Supporting Information.

***Implications of the ME effects on the topological magnon dispersion***

Although $Cu_3TeO_6$ with a centrosymmetric $Ia\overline{3}$ space group possesses a global center of inversion, Cu-O-Cu bond angles smaller than 180° lead to the breaking of the local inversion symmetry around the region where a ligand ion lies between two magnetic ions. The absence of this local inversion symmetry allows in principle the DMI leading to spin canting. According to the neutron diffraction study of Herak *et al.,* [25] the AFM ordering of $Cu_3TeO_6$ below $T_N$=62 K leads to either collinear or slightly tilted non-collinear spin configuration with a dominant magnetic moment along one of the [±1±1±1] directions. This concomitantly results in a decrease in crystal symmetry from cubic to trigonal.

According to theoretical calculations by Li *et al.*, [30] who refer to the experimental spin canting angles of 6° within the spin pair connected by the strongest interaction $J_1$, the value of $|D_1|/|J_1|$ in $Cu_3TeO_6$ was estimated as ~ 0.2. However, Wang *et al.* [33] has considered all the $J_i$ (i=1-20) superexchange interactions, in which $J_9$ is the second strongest one, and estimated $|D_1|/|J_1|$ ~0.06 in $Cu_3TeO_6$, resulting in only the spin canting angle of only ~1.3° between the $J_1$ pair. Moreover, in this case, the value of $|D_9|$ corresponding to the second strongest $J_9$ was found to be negligible, because the bond angle corresponding to this exchange coupling (i.e., $J_9$) is very close to 180°. As a result, Wang *et al.* [33] have predicted that the length of the nodal line should be proportional to $[|D_i|/|J_i|]^2$. However, the expected length of the nodal line is indeed very short so that it is difficult to differentiate the nodal line from the Dirac point according to the available INS data. [31,32]

According to the theoretical calculations of Li *et al.* [30] and Wang *et al.,* [33] the Dirac point and the Nodal line in the magnon dispersion are coined to the absence and the presence of a spin canting in the AFM state of $Cu_3TeO_6$, respectively. However, from our point of view, the prediction of nodal lines in $Cu_3TeO_6$ is the consequence of the simplicity of the model with only two exchange interactions. [30] Indeed, experimental INS results suggest the absence of nodal lines, and an involvement of many exchange paths, which is also supported by theoretical calculations. [31,32] The INS experiment, however, indicates possible presence of Dirac points in the Γ and P points of the Brillouin zone, which should be allowed if the conservation of $I \otimes t$ -symmetry is valid.

Below $T_N$, the linear ME effect causes the electric and magnetic fields to have the same symmetry in $Cu_3TeO_6$. As a result, applying either field in $Cu_3TeO_6$ will break the remaining $I \otimes t$ symmetry, leading to either the transformation of Dirac points into the Weyl points or the appearance of energy gaps. The extent of the splitting or gap formation will be proportional to the applied field strength. Since magnons are neutral without any orbital effect and do not change quasi-momentum under external fields, the Weyl points will persist even under strong fields. Therefore, the magnon spectra in $Cu_3TeO_6$ can be manipulated by both $H$ and electric field via a Zeeman coupling and the ME coupling, respectively. We expect both magnetic and electric fields to be equally effective in manipulating the topological magnon structure, but experimental verification is required to confirm their effectiveness.

## CONCLUSION

In conclusion, we observed a linear ME effect in $Cu_3TeO_6$ with a maximum value of $P$ = 20 μC/m$^2$ at $\mu_0 H$ = 14 T and a corresponding linear ME coefficient $\alpha$ = 1.8 ps/m for the $P$ // [110] and $H$ // [1-10] configuration. Moreover, the ME coefficient is one order of magnitude higher for transversal configurations compared to parallel configurations. Detailed symmetry analysis indicates that the linear ME effect originates from the ME tensor components allowed by the magnetic point group $\bar{3}'$ symmetry. Thus, $Cu_3TeO_6$ is a rare material that exhibits both topological quantum spin excitations and linear magnetoelectricity, demonstrating the richness of its physics. We suggest that the manipulation of topological properties can be achieved by both applied electric and magnetic fields, leading to the transformation of Dirac points into Weyl points or the appearance of energy gaps. This work will pave the way for electric field control of topological magnon dispersion in $Cu_3TeO_6$. Further research on neutron scattering under magnetic or electric fields, as well as theoretical simulation, will be useful in shedding more light on the intriguing possibility of manipulating topological magnon dispersion.

## ACKNOWLEDGMENTS

This work was supported by the Ministry of Science and ICT through NRF of the National Research Foundation of South Korea funded by Ministry of Science and ICT (2019R1A2C2090648, 2022H1D3A3A01077468) and core facility program funded by the Ministry of Education (2021R1A6C101B418). N.V.T. acknowledges financial support by the Ministry of Science and Higher Education of the Russian Federation (State assignment in the field of scientific activity, Southern Federal University, 2023, Project No. FENW-2023-0015) and A.S. acknowledges financial support from the Department of Science and Technology in India through the Ramanujan Faculty Award (RJF/2021/000122/SSC).

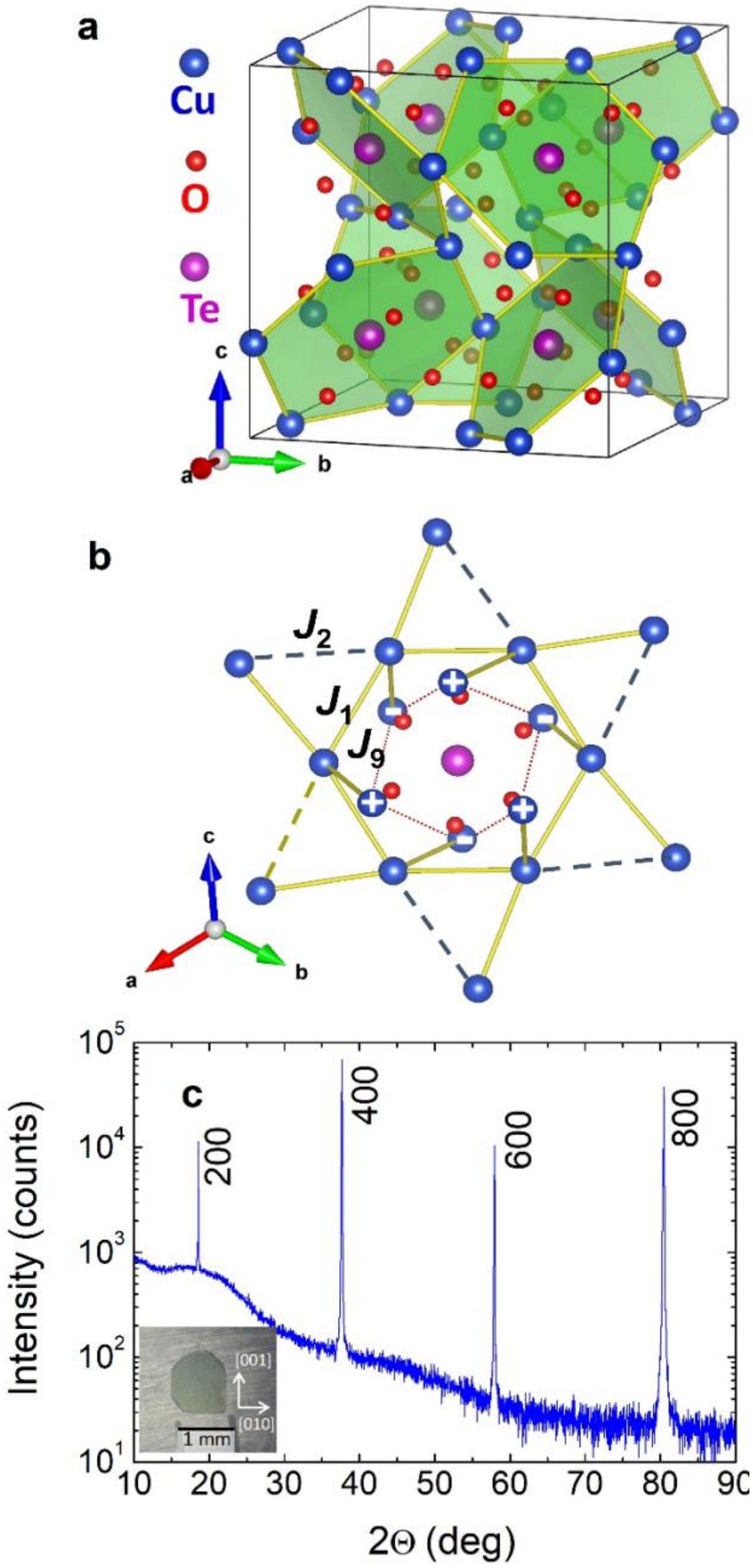

Figure 1 (a) The crystal structure of $Cu_3TeO_6$ including the eight Cu hexagons surrounding $TeO_6$ octahedra. (b) First (solid line), second (bold dash line), and nineth (weak dash line) nearest neighbor (NN) exchange interactions are drawn as $J_1$, $J_2$ , and $J_9$, respectively. The plus (+) and minus (-) signs indicate Cu ions of neighbouring Cu hexagon at equal distance above and below the shown Cu hexagon of solid line and $J_9$ is exchange interactions between these two Cu ions. The 1$^{st}$ NN forms a hexagon and the 1$^{st}$ with the 2$^{nd}$ NN forms a hyper kagome structure. (c) XRD data taken along the [100] crystallographic direction with the inset picture of the $Cu_3TeO_6$ single crystal.

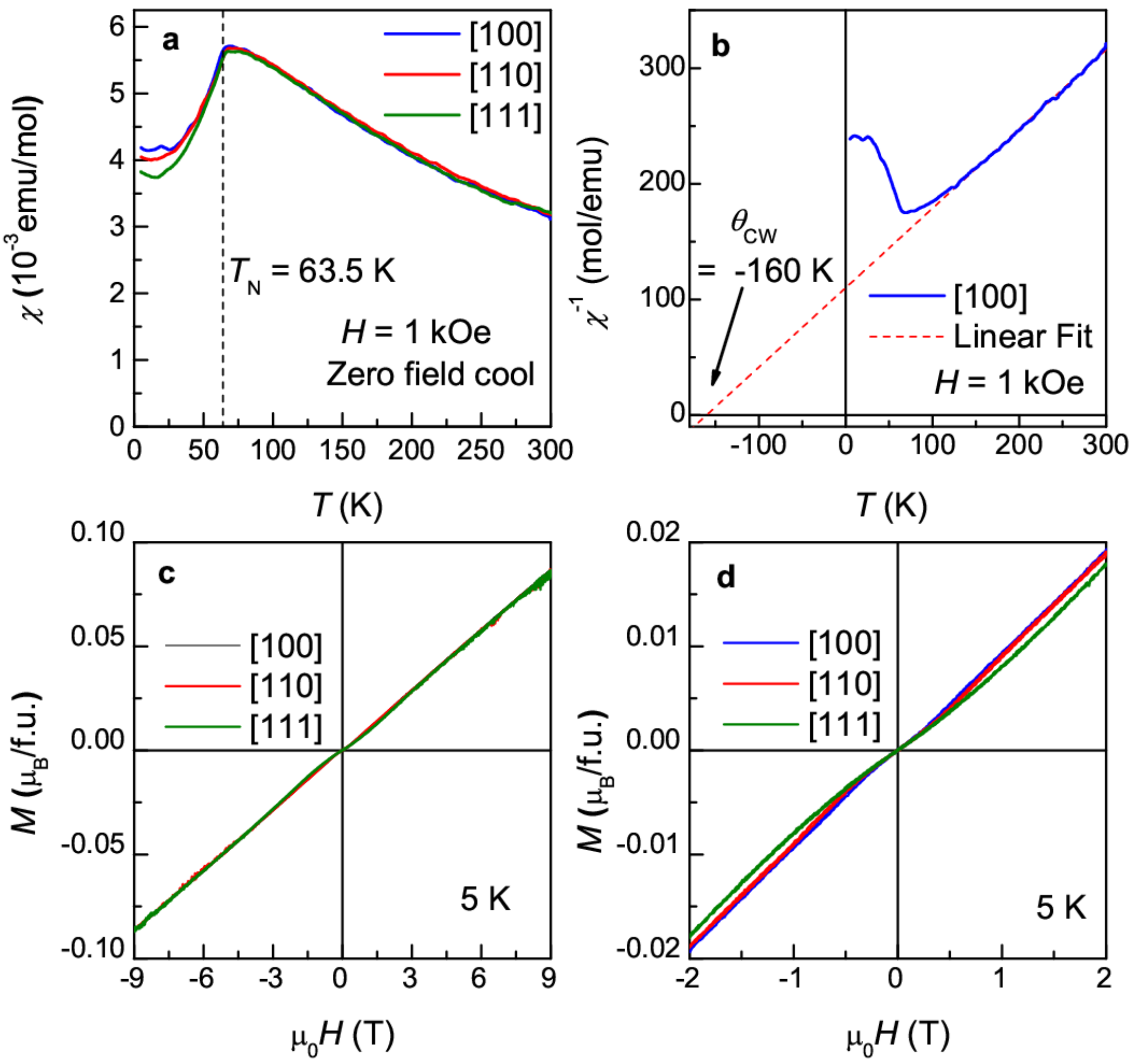

Figure 2 Temperature and magnetic dependence of magnetic properties of $Cu_3TeO_6$. (a) Magnetic susceptibility and (b) inverse susceptibility as a function of temperature and (c,d) magnetic field dependence of magnetization with two field ranges along each direction in the $Cu_3TeO_6$ single crystals .

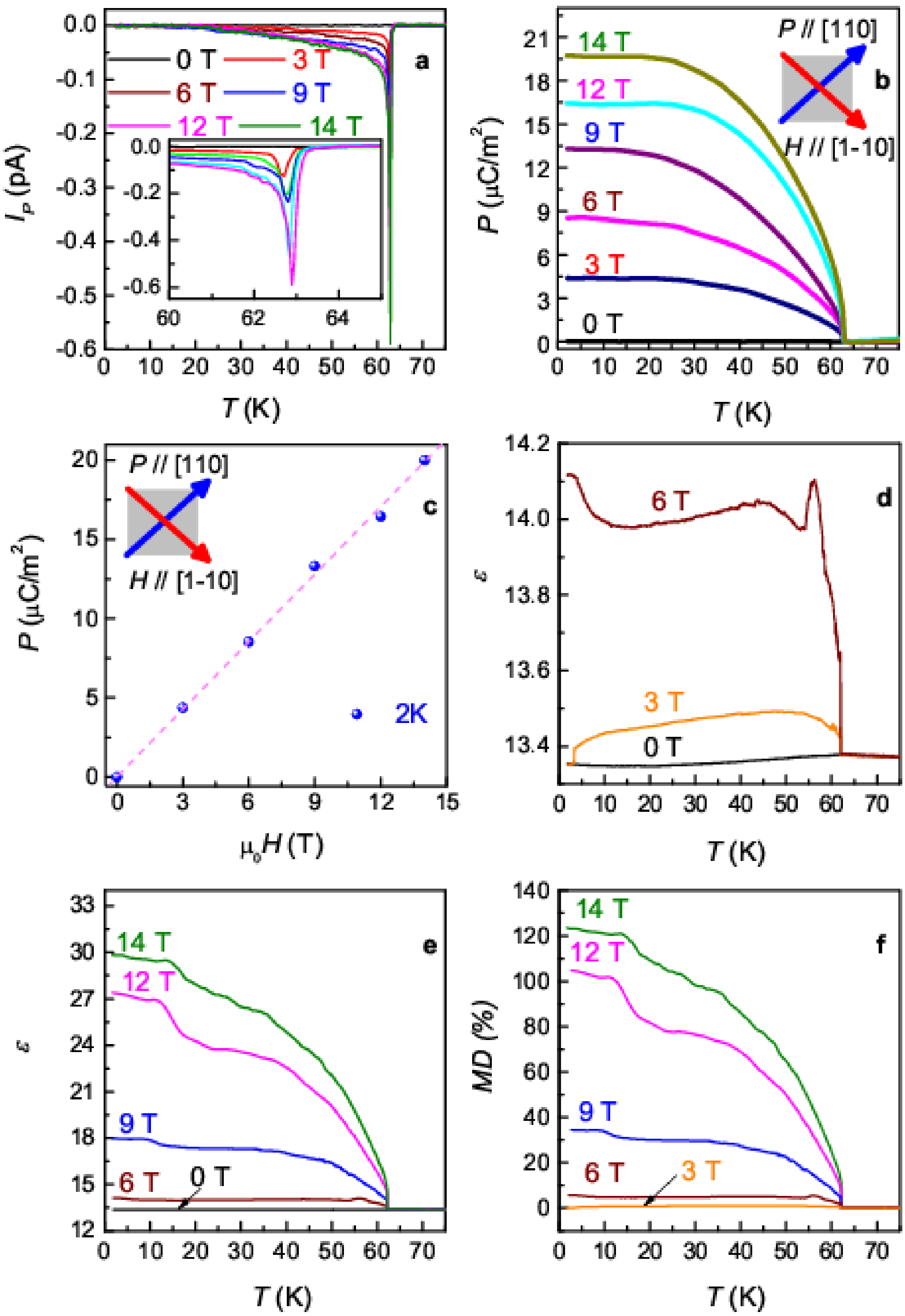

Figure 3 Temperature dependence of electric properties of $Cu_3TeO_6$ . The (a) pyroelectric current along [110] for *E*//[110] at various *H*//[1-10], (b) the electric polarization *P*//[110] at various *H*//[1-10], (c) summarizes the electric polarization *P*//[110] at various *H*//[1-10] at 2 K. (d,e) the dielectric constant for *P*,*E*//[110] and *H*//[1-10] and (f) the corresponding magnetodielectric (MD) effect in $Cu_3TeO_6$ single crystals.

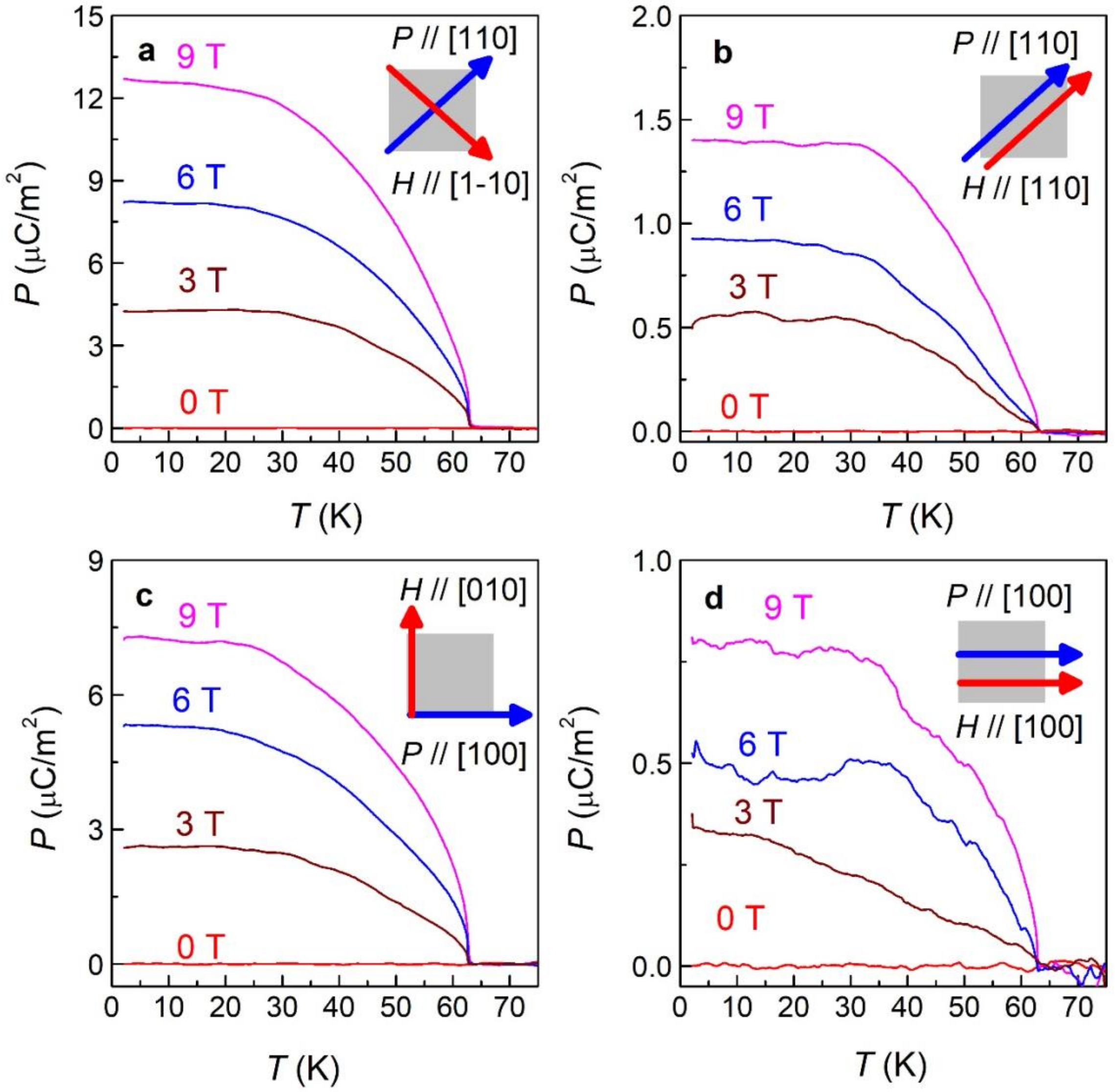

Figure 4 Temperature dependence of electric polarization at selected magnetic fields in the $Cu_3TeO_6$ single crystals for various *P* and *H* directions as represented by schematic inset figures for each data. The polarization becomes significantly large in the transverse configurations, supporting the presence of the dominant off-diagonal ME tensors in $Cu_3TeO_6$.

## REFERENCES

[1] D. N. Astrov, *The Magnetoelectric Effect in Antiferromagnetics*, Journal of Experimental and Theoretical Physics **38**, 984 (1960).

[2] L. D. Landau and E. M. Lifshitz, *Ferromagnetism and Antiferromagnetism*, in *Electrodynamics of Continuous Media*, Vol. 29 (Elsevier, 1984), pp. 130–179.

[3] I. E. Dzyaloshinskii, *On the Magnetoelectrical Effect in Antiferromagnetics*, Journal of Experimental and Theoretical Physics **37**, 881 (1959).

[4] T. H. T. O'Dell, *The Electrodynamics of Magneto-Electric Media* (North-Holland, Amsterdam, 1970).

[5] G. A. Smolenskiĭ and I. E. Chupis, *Ferroelectromagnets*, Soviet Physics Uspekhi **25**, 475 (1982).

[6] J. Wang et al., *Epitaxial BiFeO3 Multiferroic Thin Film Heterostructures*, Science (1979) **299**, 1719 (2003).

[7] T. Kimura, T. Goto, H. Shintani, K. Ishizaka, T. Arima, and Y. Tokura, *Magnetic Control of Ferroelectric Polarization*, Nature 2003 426:6962 **426**, 55 (2003).

[8] N. Hur, S. Park, P. A. Sharma, J. S. Ahn, S. Guha, and S. W. Cheong, *Electric Polarization Reversal and Memory in a Multiferroic Material Induced by Magnetic Fields*, Nature 2004 429:6990 **429**, 392 (2004).

[9] S.-W. Cheong and M. Mostovoy, *Multiferroics: A Magnetic Twist for Ferroelectricity*, Nat Mater **6**, 13 (2007).

[10] N. A. Spaldin and R. Ramesh, *Advances in Magnetoelectric Multiferroics*, Nature Materials 2019 18:3 **18**, 203 (2019).

[11] C. B. Park, A. Shahee, K. Kim, D. R. Patil, S. A. Guda, N. Ter-Oganessian, and K. H. Kim, *Observation of Spin-Induced Ferroelectricity in a Layered van Der Waals Antiferromagnet CuCrP2S6*, Adv Electron Mater **8**, 2101072 (2022).

[12] L. Ponet, S. Artyukhin, T. Kain, J. Wettstein, A. Pimenov, A. Shuvaev, X. Wang, S. W. Cheong, M. Mostovoy, and A. Pimenov, *Topologically Protected Magnetoelectric Switching in a Multiferroic*, Nature 2022 607:7917 **607**, 81 (2022).

[13] K. W. Shin, M. Soroka, A. Shahee, K. H. Kim, J. Buršík, R. Kužel, M. Vronka, and M. H. Aguirre, *Observation of Anomalously Large Magnetoelectric Coupling in the Hexagonal Z-Type Ferrite Films*, Adv Electron Mater **8**, 2101294 (2022).

[14] J. Zhai, Z. Xing, S. Dong, J. Li, and D. Viehland, *Detection of Pico-Tesla Magnetic Fields Using Magneto-Electric Sensors at Room Temperature*, Appl Phys Lett **88**, 062510 (2006).

[15] M. Bibes and A. Barthélémy, *Towards a Magnetoelectric Memory*, Nat. Mater. **7**, 425 (2008).

[16] Y. Li and K.-M. Luk, *A Multibeam End-Fire Magnetoelectric Dipole Antenna Array for Millimeter-Wave Applications*, IEEE Trans Antennas Propag **64**, 2894 (2016).

[17] S. Manipatruni, D. E. Nikonov, C. Lin, tanay A. Gosavi, H. Liu, B. Prasad, Y. Huang, E.

Bonturim, R. Ramesh, and ian A. Young, *Scalable Energy-Efficient Magnetoelectric Spin–Orbit Logic*, Nature **565**, 35 (2019).

[18] S. K. Ghosh, K. Roy, H. K. Mishra, M. R. Sahoo, B. Mahanty, P. N. Vishwakarma, and D. Mandal, *Rollable Magnetoelectric Energy Harvester as a Wireless IoT Sensor*, ACS Sustain Chem Eng **8**, 864 (2020).

[19] S. A. Ivanov, C. Ritter, P. Nordblad, R. Tellgren, M. Weil, V. Carolus, T. Lottermoser, M. Fiebig, and R. Mathieu, *New Insights into the Multiferroic Properties of Mn3TeO6*, J Phys D Appl Phys **50**, 85001 (2017).

[20] L. Zhao, Z. Hu, C. Y. Kuo, T. W. Pi, M. K. Wu, L. H. Tjeng, and A. C. Komarek, *Mn3TeO6 - a New Multiferroic Material with Two Magnetic Substructures*, Physica Status Solidi - Rapid Research Letters **9**, 730 (2015).

[21] M. Hudl et al., *Complex Magnetism and Magnetic-Field-Driven Electrical Polarization of Co3TeO6*, Phys Rev B **84**, 180404 (2011).

[22] N. Leo et al., *Magnetoelectric Inversion of Domain Patterns*, Nature **560**, 466 (2018).

[23] Y. S. Oh, S. Artyukhin, J. J. Yang, V. Zapf, J. W. Kim, D. Vanderbilt, and S. W. Cheong, *Non-Hysteretic Colossal Magnetoelectricity in a Collinear Antiferromagnet*, Nat Commun **5**, 3201 (2014).

[24] J. W. Kim et al., *Successive Magnetic-Field-Induced Transitions and Colossal Magnetoelectric Effect in Ni3TeO6*, Phys Rev Lett **115**, 137201 (2015).

[25] M. Herak, H. Berger, M. Prester, M. Miljak, I. Živković, O. Milat, D. Drobac, S. Popović, and O. Zaharko, *Novel Spin Lattice in Cu3TeO6: An Antiferromagnetic Order and Domain Dynamics*, Journal of Physics Condensed Matter **17**, 7667 (2005).

[26] L. Falck, O. Lindqvist, and J. Moret, *Tricopper(II) Tellurate(VI)*, Acta Crystallogr B **34**, 896 (1978).

[27] M. Herak, *Cubic Magnetic Anisotropy of the Antiferromagnetically Ordered Cu3TeO6*, Solid State Commun **151**, 1588 (2011).

[28] Z. He and M. Itoh, *Magnetic Behaviors of Cu3TeO6 with Multiple Spin Lattices*, J Magn Magn Mater **354**, 146 (2014).

[29] M. Månsson, K. Prša, J. Sugiyama, D. Andreica, H. Luetkens, and H. Berger, *Magnetic Order and Transitions in the Spin-Web Compound Cu3TeO6*, Phys Procedia **30**, 142 (2012).

[30] K. Li, C. Li, J. Hu, Y. Li, and C. Fang, *Dirac and Nodal Line Magnons in Three-Dimensional Antiferromagnets*, Phys Rev Lett **119**, 247202 (2017).

[31] S. Bao et al., *Discovery of Coexisting Dirac and Triply Degenerate Magnons in a Three-Dimensional Antiferromagnet*, Nat Commun **9**, 2591 (2018).

[32] W. Yao, C. Li, L. Wang, S. Xue, Y. Dan, K. Iida, K. Kamazawa, K. Li, C. Fang, and Y. Li, *Topological Spin Excitations in a Three-Dimensional Antiferromagnet*, Nat Phys **14**, 1011 (2018).

[33] D. Wang, X. Bo, F. Tang, and X. Wan, *Calculated Magnetic Exchange Interactions in the Dirac Magnon Material Cu3TeO6*, Phys Rev B **99**, 035160 (2019).

[34] S. A. Owerre, *Floquet Weyl Magnons in Three-Dimensional Quantum Magnets*, Sci Rep **8**, 10098 (2018).

[35] K. Y. Choi, P. Lemmens, E. S. Choi, and H. Berger, *Lattice Anomalies and Magnetic Excitations of the Spin Web Compound $Cu^3TeO^6$*, Journal of Physics Condensed Matter **20**, 505214 (2008).

[36] G. Caimi, L. Degiorgi, H. Berger, and L. Forró, *Optical Evidence for a Magnetically Driven Structural Transition in the Spin Web Cu3TeO6*, Europhys Lett **75**, 496 (2006).

[37] J. P. Rivera, *A Short Review of the Magnetoelectric Effect and Related Experimental Techniques on Single Phase (Multi-) Ferroics*, The European Physical Journal B 2009 71:3 **71**, 299 (2009).

[38] J.-P. Rivera and H. Schmid, *Search for the Piezomagnetoelectric Effect in LiCoPO4*, Ferroelectrics **161**, 91 (1994).

[39] H. Niu et al., *Room Temperature Magnetically Ordered Polar Corundum GaFeO3 Displaying Magnetoelectric Coupling*, J Am Chem Soc **139**, 1520 (2017).

[40] K. D. Hughey et al., *High-Field Magnetoelectric and Spin-Phonon Coupling in Multiferroic (NH4)2[FeCl5·(H 2O)]*, Inorg Chem **61**, 3434 (2022).

[41] D. Vaknin, J. L. Zarestky, J.-P. Rivera, and H. Schmid, *Commensurate-Incommensurate Magnetic Phase Transition in Magnetoelectric Single Crystal LiNiPO4*, Phys Rev Lett **92**, 207201 (2004).

[42] A. Iyama and T. Kimura, *Magnetoelectric Hysteresis Loops in Cr2O3 at Room Temperature*, Phys Rev B **87**, 180408 (2013).

[43] X. Li et al., *Magnetoelectric Mutual-Control in Collinear Antiferromagnetic NdCrTiO5*, Appl Phys Lett **113**, (2018).

[44] P. Yanda, N. V. Ter-Oganessian, and A. Sundaresan, *Linear Magnetoelectric Effect in Antiferromagnetic Sm2BaCuO5*, Phys Rev B **100**, 104417 (2019).

[45] K. Yoo et al., *Magnetic Field-Induced Ferroelectricity in S = 1/2 Kagome Staircase Compound PbCu3TeO7*, NPJ Quantum Mater **3**, 45 (2018).

## Supporting Information

### Magnetic basis vectors

The cubic structure of $Cu_3TeO_6$ with space group $Ia\overline{3}$ contains twelve $Cu^{2+}$ ions in the primitive cell whose positions are given in Table S1. Assuming that $\vec{S}_i$ is the spin of ion i one can find the basis vectors for magnetic structures with $\vec{k} = 0$ in $Cu_3TeO_6$, which are summarized in Table S2.

Table S1. Positions of $Cu^{2+}$ ions in $Cu_3TeO_6$.

| Atom | Position | | |
|---|---|---|---|
| | x | y | z |
| $Cu_1$ | 0.969 | 0 | $1/4$ |
| $Cu_2$ | 0 | $1/4$ | 0.969 |
| $Cu_3$ | $1/4$ | 0.969 | 0 |
| $Cu_4$ | 0.531 | 0 | $3/4$ |
| $Cu_5$ | 0 | $3/4$ | 0.531 |
| $Cu_6$ | $3/4$ | 0.531 | 0 |
| $Cu_7$ | 0.031 | 0 | $3/4$ |
| $Cu_8$ | 0 | $3/4$ | 0.031 |
| $Cu_9$ | $3/4$ | 0.031 | 0 |
| $Cu_{10}$ | 0.469 | 0 | $1/4$ |
| $Cu_{11}$ | 0 | $1/4$ | 0.469 |
| $Cu_{12}$ | $1/4$ | 0.469 | 0 |

Table S2. Basis vectors for magnetic structures with $\vec{k} = 0$ in $Cu_3TeO_6$.

| **IR** | **Basis vectors** |
|---|---|
| $\Gamma^{1+}$ | $S_{1x} - S_{10x} - S_{4x} + S_{7x} - S_{12y} + S_{3y} - S_{6y} + S_{9y} - S_{11z} + S_{2z} - S_{5z} + S_{8z}$ |
| $\Gamma^{1-}$ | $S_{1x} + S_{10x} - S_{4x} - S_{7x} + S_{12y} + S_{3y} - S_{6y} - S_{9y} + S_{11z} + S_{2z} - S_{5z} - S_{8z}$ |
| $\Gamma^{2+3+}$ | $\begin{pmatrix} 2S_{1x} - 2S_{10x} - 2S_{4x} + 2S_{7x} + S_{12y} - S_{3y} + S_{6y} - S_{9y} + S_{11z} - S_{2z} + S_{5z} - S_{8z} \\ \sqrt{3}S_{12y} - \sqrt{3}S_{3y} + \sqrt{3}S_{6y} - \sqrt{3}S_{9y} - \sqrt{3}S_{11z} + \sqrt{3}S_{2z} - \sqrt{3}S_{5z} + \sqrt{3}S_{8z} \end{pmatrix}$ |
| $\Gamma^{2-3-}$ | $\begin{pmatrix} 2S_{1x} + 2S_{10x} - 2S_{4x} - 2S_{7x} - S_{12y} - S_{3y} + S_{6y} + S_{9y} - S_{11z} - S_{2z} + S_{5z} + S_{8z} \\ -\sqrt{3}S_{12y} - \sqrt{3}S_{3y} + \sqrt{3}S_{6y} + \sqrt{3}S_{9y} + \sqrt{3}S_{11z} + \sqrt{3}S_{2z} - \sqrt{3}S_{5z} - \sqrt{3}S_{8z} \end{pmatrix}$ |
| $\Gamma^{4+}$ | $\begin{pmatrix} S_{1x} + S_{10x} + S_{4x} + S_{7x} \\ S_{12y} + S_{3y} + S_{6y} + S_{9y} \\ S_{11z} + S_{2z} + S_{5z} + S_{8z} \end{pmatrix}$ |
| $\Gamma^{4+}$ | $\begin{pmatrix} S_{11x} + S_{2x} + S_{5x} + S_{8x} \\ S_{1y} + S_{10y} + S_{4y} + S_{7y} \\ S_{12z} + S_{3z} + S_{6z} + S_{9z} \end{pmatrix}$ |
| $\Gamma^{4+}$ | $\begin{pmatrix} -S_{11y} + S_{2y} - S_{5y} + S_{8y} \\ S_{1z} - S_{10z} - S_{4z} + S_{7z} \\ -S_{12x} + S_{3x} - S_{6x} + S_{9x} \end{pmatrix}$ |
| $\Gamma^{4+}$ | $\begin{pmatrix} S_{12x} + S_{3x} + S_{6x} + S_{9x} \\ S_{11y} + S_{2y} + S_{5y} + S_{8y} \\ S_{1z} + S_{10z} + S_{4z} + S_{7z} \end{pmatrix}$ |
| $\Gamma^{4+}$ | $\begin{pmatrix} -S_{12z} + S_{3z} - S_{6z} + S_{9z} \\ -S_{11x} + S_{2x} - S_{5x} + S_{8x} \\ S_{1y} - S_{10y} - S_{4y} + S_{7y} \end{pmatrix}$ |
| $\Gamma^{4-}$ | $\begin{pmatrix} S_{1x} - S_{10x} + S_{4x} - S_{7x} \\ -S_{12y} + S_{3y} + S_{6y} - S_{9y} \\ -S_{11z} + S_{2z} + S_{5z} - S_{8z} \end{pmatrix}$ |
| $\Gamma^{4-}$ | $\begin{pmatrix} -S_{11x} + S_{2x} + S_{5x} - S_{8x} \\ S_{1y} - S_{10y} + S_{4y} - S_{7y} \\ -S_{12z} + S_{3z} + S_{6z} - S_{9z} \end{pmatrix}$ |
| $\Gamma^{4-}$ | $\begin{pmatrix} S_{11y} + S_{2y} - S_{5y} - S_{8y} \\ S_{1z} + S_{10z} - S_{4z} - S_{7z} \\ S_{12x} + S_{3x} - S_{6x} - S_{9x} \end{pmatrix}$ |
| $\Gamma^{4-}$ | $\begin{pmatrix} -S_{12x} + S_{3x} + S_{6x} - S_{9x} \\ -S_{11y} + S_{2y} + S_{5y} - S_{8y} \\ S_{1z} - S_{10z} + S_{4z} - S_{7z} \end{pmatrix}$ |
| $\Gamma^{4-}$ | $\begin{pmatrix} S_{12z} + S_{3z} - S_{6z} - S_{9z} \\ S_{11x} + S_{2x} - S_{5x} - S_{8x} \\ S_{1y} + S_{10y} - S_{4y} - S_{7y} \end{pmatrix}$ |

**Magnetoelectric interactions**

The ME interactions (1) and (2) can be written in the form

$$P_x\left((S_{3y}+S_{6y})(S_{3z}+S_{6z})-(S_{12y}+S_{9y})(S_{12z}+S_{9z})\right) \\ +P_y\left((S_{2x}+S_{5x})(S_{2z}+S_{5z})-(S_{11x}+S_{8x})(S_{11z}+S_{8z})\right) \\ +P_z\left((S_{1x}+S_{4x})(S_{1y}+S_{4y})-(S_{10x}+S_{7x})(S_{10y}+S_{7y})\right),$$

$$P_x\left((S_{2y}+S_{5y})(S_{2z}+S_{5z})-(S_{11y}+S_{8y})(S_{11z}+S_{8z})\right) \\ +P_y\left((S_{1x}+S_{4x})(S_{1z}+S_{4z})-(S_{10x}+S_{7x})(S_{10z}+S_{7z})\right) \\ +P_z\left((S_{3x}+S_{6x})(S_{3y}+S_{6y})-(S_{12x}+S_{9x})(S_{12y}+S_{9y})\right),$$

$$P_x\left((S_{3y}+S_{6y})(S_{3z}+S_{6z})-(S_{12y}+S_{9y})(S_{12z}+S_{9z})\right) \\ +P_y\left((S_{2x}+S_{5x})(S_{2z}+S_{5z})-(S_{11x}+S_{8x})(S_{11z}+S_{8z})\right) \\ +P_z\left((S_{1x}+S_{4x})(S_{1y}+S_{4y})-(S_{10x}+S_{7x})(S_{10y}+S_{7y})\right),$$

$$P_x\left((S_{3y}+S_{6y})(S_{10z}+S_{7z})-(S_{12y}+S_{9y})(S_{1z}+S_{4z})\right) \\ +P_y\left((S_{12x}+S_{9x})(S_{2z}+S_{5z})-(S_{3x}+S_{6x})(S_{11z}+S_{8z})\right) \\ +P_z\left((S_{1x}+S_{4x})(S_{11y}+S_{8y})-(S_{10x}+S_{7x})(S_{2y}+S_{5y})\right),$$

$$P_x\left((S_{3y}+S_{6y})(S_{11z}+S_{8z})-(S_{12y}+S_{9y})(S_{2z}+S_{5z})\right) \\ +P_y\left((S_{10x}+S_{7x})(S_{2z}+S_{5z})-(S_{1x}+S_{4x})(S_{11z}+S_{8z})\right) \\ +P_z\left((S_{1x}+S_{4x})(S_{12y}+S_{9y})-(S_{10x}+S_{7x})(S_{3y}+S_{6y})\right),$$

$$P_x\left((S_{1y}+S_{4y})(S_{12z}+S_{9z})-(S_{10y}+S_{7y})(S_{3z}+S_{6z})\right) \\ +P_y\left((S_{11x}+S_{8x})(S_{3z}+S_{6z})-(S_{2x}+S_{5x})(S_{12z}+S_{9z})\right) \\ +P_z\left((S_{2x}+S_{5x})(S_{10y}+S_{7y})-(S_{11x}+S_{8x})(S_{1y}+S_{4y})\right),$$

$$P_x\left((S_{1y}+S_{4y})(S_{11z}+S_{8z})-(S_{10y}+S_{7y})(S_{2z}+S_{5z})\right)$$
$$+P_y\left((S_{10x}+S_{7x})(S_{3z}+S_{6z})-(S_{1x}+S_{4x})(S_{12z}+S_{9z})\right)$$
$$+P_z\left((S_{2x}+S_{5x})(S_{12y}+S_{9y})-(S_{11x}+S_{8x})(S_{3y}+S_{6y})\right),$$
$$P_x\left((S_{2y}+S_{5y})(S_{12z}+S_{9z})-(S_{11y}+S_{8y})(S_{3z}+S_{6z})\right)$$
$$+P_y\left((S_{11x}+S_{8x})(S_{1z}+S_{4z})-(S_{2x}+S_{5x})(S_{10z}+S_{7z})\right)$$
$$+P_z\left((S_{3x}+S_{6x})(S_{10y}+S_{7y})-(S_{12x}+S_{9x})(S_{1y}+S_{4y})\right),$$
$$P_x\left((S_{2y}+S_{5y})(S_{10z}+S_{7z})-(S_{11y}+S_{8y})(S_{1z}+S_{4z})\right)$$
$$+P_y\left((S_{12x}+S_{9x})(S_{1z}+S_{4z})-(S_{3x}+S_{6x})(S_{10z}+S_{7z})\right)$$
$$+P_z\left((S_{3x}+S_{6x})(S_{11y}+S_{8y})-(S_{12x}+S_{9x})(S_{2y}+S_{5y})\right).$$

Here we consider only single ion (the first three equations contain single ion contributions) and nearest neighbor (remaining equations) interactions. One can see that in fact the ME interactions are numerous and difficult to analyze.

## Monte Carlo calculations

Several sets of magnetic exchange constants are available for $Cu_3TeO_6$, that were obtained by analysis of inelastic neutron diffraction or density functional calculations. We performed Monte Carlo (MC) calculations with published magnetic exchange constants using the Hamiltonian

$$H=\sum_{i,j} J_{ij}\vec{S}_i\vec{S}_j, \qquad (1)$$

where the summation runs over all *ij* pairs of spins with exchange constants $J_{ij}$, and $\vec{S}_i$ are unit vectors. The Metropolis scheme was used for MC calculations of the system with dimensions 10×10×10 cubic cells each containing 24 spins. After each temperature change the system was relaxed for $3\cdot10^3$ Monte Carlo steps per spin (MCS), whereas the statistical information was gathered over the next $3\cdot10^3$ MCS. Table S3 summarizes the magnetic exchange constants obtained by inelastic neutron diffraction (Sets 1, 2, 3) [1] and density functional calculations (Set 4) [2], which were used in MC calculations. The latter set of exchange parameters is that with *U*=10 eV in the LSDA+*U* calculations utilized in Ref. [2].

Table S3. Magnetic exchange constants in $Cu_3TeO_6$.

| $J$ | Distance, Å | Exchange energy, meV | | | |
|---|---|---|---|---|---|
| | | Set 1 [1] | Set 2 [1] | Set 3 [1] | Set 4 [2] |
| $J_1$ | 3.18 | 4.5 | 4.4 | 4.4 | 7.05 |
| $J_2$ | 3.60 | -0.22 | -0.36 | -0.41 | 0.51 |
| $J_3$ | 4.77 | -1.5 | -1.6 | -1.63 | 0.04 |
| $J_4$ | 4.81 | 1.33 | 1.3 | 1.31 | 2.18 |
| $J_5$ | 4.81 | 1.8 | 1.5 | 1.7 | 0.09 |
| $J_6$ | 5.48 | -0.21 | -0.21 | -0.21 | 0.01 |
| $J_7$ | 5.73 | -0.14 | -0.2 | -0.14 | -0.01 |
| $J_8$ | 5.97 | 0.11 | 0.03 | 0.054 | 0.04 |
| $J_9$ | 6.21 | 4.5 | 4.5 | 4.26 | 3.77 |
| $J_{10}$ | 6.34 | | | | 0.56 |
| $J_{11}$ | 6.34 | | | | -0.01 |
| $J_{12}$ | 6.74 | | | | 0.02 |
| $J_{13}$ | 7.17 | | | | -0.06 |
| $J_{14}$ | 7.17 | | | | -0.04 |
| $J_{15}$ | 7.27 | | | | 0 |
| $J_{16}$ | 7.46 | | | | 0.02 |
| $J_{17}$ | 7.64 | | | | 0.1 |
| $J_{18}$ | 7.83 | | | | 0 |
| $J_{19}$ | 8.26 | | | | -0.02 |
| $J_{20}$ | 8.26 | | | | 0 |

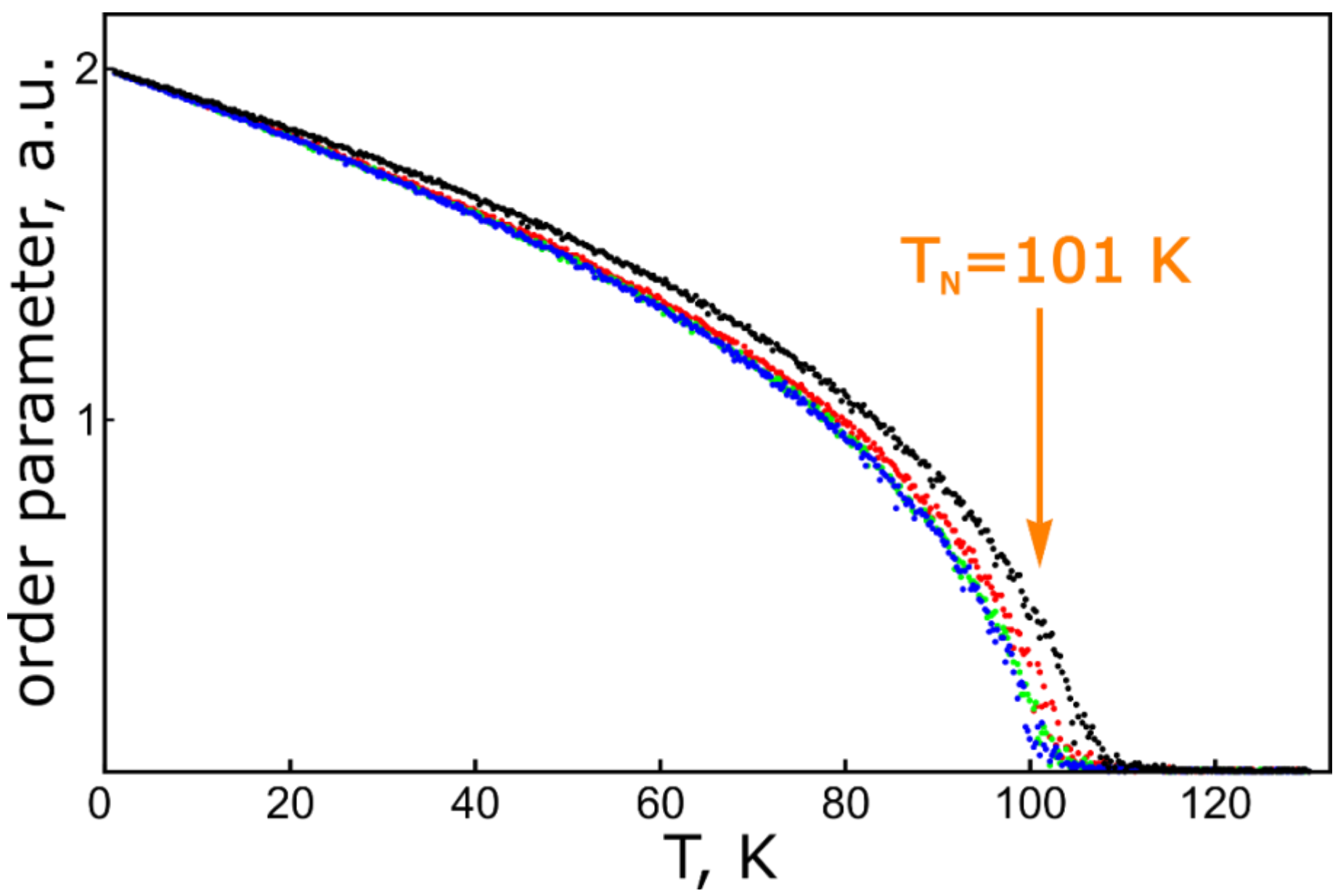

Figure S1. Temperature dependence of antiferromagnetic order parameter in MC calculations using the magnetic exchange sets 1 (red), 2 (green), 3 (blue), and 4 (black).

Figure S1 shows the temperature dependence of the absolute value of antiferromagnetic order parameter $(a_1, a_2, a_3)$. It can be found that all sets of exchange parameters give the antiferromagnetic ordering corresponding to experimental results, whereas the Neel temperatures are in the range ~101 -108 K in disagreement with the experimental value of 61 K. It has to be noted, though, that this quantitative discrepancy of $T_N$ can be due to different definitions of exchange constants, because in our model we use the Hamiltonian (1) with unit spins.

**References**

1. W. Yao, C. Li, L. Wang, S. Xue, Y. Dan, K. Iida, K. Kamazawa, K. Li, C. Fang, Y. Li, Topological spin excitations in a three-dimensional antiferromagnet, Nat. Phys. **14**, 1011 (2018).
2. D. Wang, X. Bo, F. Tang, X. Wan, Calculated magnetic exchange interactions in the Dirac magnon material $Cu_3TeO_6$, Phys. Rev. B **99**, 035160 (2019).